\documentclass[11pt]{scrartcl}
\pdfoutput=1
\usepackage{listings}
\usepackage{amsmath}
\usepackage{amssymb}
\usepackage{amsthm}
\usepackage{tikz}
\usepackage{mathpartir}
\usepackage{braket}
\usepackage{hyperref}

\theoremstyle{definition}
\newtheorem{example}{Example}

\theoremstyle{plain}
\newtheorem{definition}{Definition}

\title{Binary-Decision-Diagrams for Set Abstraction}

\author{Arlen Cox\\University of Colorado Boulder/ENS Paris}

\date{}

\begin{document}
\newcommand{\universe}{\ensuremath{\mathbb{U}}}
\newcommand{\comp}[1]{\ensuremath{{#1}^{\mathrm{c}}}}
\newcommand{\itef}[1]{\ensuremath{\mathrm{ite}({#1})}}
\newcommand{\ite}[3]{\itef{{#1},{#2},{#3})}}
\newcommand{\assignment}{\ensuremath{I}}
\newcommand{\variable}{\ensuremath{v}}
\newcommand{\variables}{\ensuremath{\textsc{Vars}}}
\newcommand{\powerset}[1]{\ensuremath{\mathcal{P}\left({#1}\right)}}
\newcommand{\values}{\ensuremath{\textsc{Vals}}}
\newcommand{\valuation}{\ensuremath{\eta}}
\newcommand{\conc}{\ensuremath{\textsc{Conc}}}
\newcommand{\tre}[1]{\ensuremath{\mathrm{tr}_E({#1})}}
\newcommand{\trk}[1]{\ensuremath{\mathrm{tr}_K({#1})}}
\newcommand{\letin}[1]{\ensuremath{\mathbf{let} \ {#1} \ \mathbf{in}}}

\maketitle

\begin{abstract}
  Whether explicit or implicit, sets are a critical part of many pieces of
  software.  As a result, it is necessary to develop abstractions of sets for
  the purposes of abstract interpretation, model checking, and deductive
  verification.  However, the construction of effective abstractions for sets
  is challenging because they are a higher-order construct.  It is necessary to
  reason about contents of sets as well as relationships between sets.  This
  paper presents a new abstraction for sets that is based on binary decision
  diagrams.  It is optimized for precisely and efficiently representing
  relations between sets while still providing limited support for content
  reasoning.
\end{abstract}

\section{Introduction}
\label{sec:intro}

In deductive software verification, it is common to want to verify programs that manipulate sets in some way.  In some cases, this is because sets are manipulated explicitly.  For example most languages, including Python, C++, and Java, provide data structure libraries that include sets.  However, as Kuncak observed~\cite{Kuncak:Thesis:2007}, it is also often useful to use sets to represent implicit invariants of non-set data structures, such as the set of elements stored in a list.  This is a useful invariant for verifying a list membership function, for instance.

In addition to deductive verification, it is also useful to be able to automatically analyze programs that manipulate implicit and explicit sets.  This means that invariants for sets need to be automatically inferred.  To do this we assume the approach of abstract interpretation~\cite{DBLP:conf/popl/CousotC77}, as it is a general approach.  There are several works that have developed and utilized abstractions suitable for sets.  QUIC graphs~\cite{DBLP:conf/ecoop/CoxCS13} uses a hypergraph to represent set constra

Automatic analysis of sets is a well studied space.  There are abstractions that have been used for automatically analyzing Python functions that explicitly manipulate sets~\cite{DBLP:conf/ecoop/CoxCS13}.  There are also abstractions that use sets implicitly for properties of other data structures.  For instance HOO~\cite{HOO} uses sets to abstract key sets for map-like data structures and FixBag~\cite{DBLP:conf/cav/PhamTTC11} uses sets to abstract the elements of a list.  This allows automatic verification of modular specifications of certain kinds of functions.

This paper develops a new kind of abstract domain for sets.  Rather adopting a \emph{content-centric} approach that focuses on the possible contents of each set, such as if set $A = \{1, 2, 5\}$, the abstract domain presented here adopts an \emph{as-a-whole} approach, focusing on relationships between sets, such as $A \subseteq B$.  By optimizing heavily for the as-a-whole case, we find that is useful to use different data structures than those focused more on contents.  Specifically, we find that the use of binary decision diagrams is particularly efficient and useful.

This paper describes the construction of such a binary-decision-diagram-based abstract domain through the following contributions:

\begin{itemize}
    \item We introduce a binary-decision-diagram-based abstract domain for set-manipulating programs.  This domain supports common set operations for fully-relational as-a-whole reasoning.  It utilizes the canonical, reduced representation of reduced, ordered binary decision diagrams to more efficiently abstract program states involving sets than existing abstractions for sets. (Section~\ref{sec:abstraction})
    \item We use a novel encoding that conflates both logical operations with set operations into a single binary decision diagram without loss of precision.  This encoding efficiently translates set operations into binary decision diagrams. (Section~\ref{sec:domain-ops})
    \item We provide a reduction with a value domain to augment the as-a-whole capable BDD-based set domain with content-centric value reasoning. (Section~\ref{sec:contents})
\end{itemize}


\section{Preliminaries}
\label{sec:preliminaries}

In this section we will give necessary background for boolean algebras and binary decision diagrams.  We will use the fact that set constraints form a boolean algebra to create an effective, efficient set abstraction in Section~\ref{sec:abstraction}.

\subsection{Sets as Boolean Algebras}

A \emph{Boolean algebra} is bounded lattice consisting of a top element $\mathbf{1}$ and a bottom element $\mathbf{0}$.  There are three operations in a Boolean algebra: (1) meet $\wedge$, which computes the greatest lower bound of two elements in the lattice, (2) join $\vee$, which computes the least upper bound of two elements in the lattice, and (3) complement $\neg$, which relates one lattice element to another.  A Boolean algebra has the following properties for lattice elements $a$, $b$, and $c$:

\begin{center}
\noindent\begin{tabular}{lll}
  $a \vee (b \vee c) = (a \vee b) \vee c$ &
  $a \wedge (b \wedge c) = (a \wedge b) \wedge c$ &
  associativity \\
  $a \vee b = b \vee a$ &
  $a \wedge b = b \wedge a$ &
  commutativity \\
  $a \vee \mathbf{0} = a$ &
  $a \wedge \mathbf{1} = b$ &
  identity \\
  $a \vee (b \wedge c) = (a \vee b) \wedge (a \vee c)$ &
  $a \wedge (b \vee c) = (a \wedge b) \vee (a \wedge c)$ &
  distributivity \\
  $a \vee \neg a = \mathbf{1}$ &
  $a \wedge \neg a = \mathbf{0}$ &
  complements \\
\end{tabular}
\end{center}

The language of sets is also a Boolean algebra.  The universal set $\universe$ is the top element.  The empty set $\emptyset$ is the bottom element.  The intersection operation $\cap$ is the meet operation.  The union operation $\cup$ is the join operation.  Finally, the set complement operation $\comp{}$ is the complement operation.

\subsection{Binary Decision Diagrams}

Binary decision diagrams (BBDs) canonically and efficiently represent a Boolean algebra.  They are based on the if-then-else (ITE) normal form:

\begin{definition}[if-then-else normal form]
	If-then-else normal form represents a Boolean algebra with the following syntactic structure:
	\begin{align*}
	  \begin{array}{rcl}
	  B & ::= & \ite{\variable}{B_t}{B_e} \\
	    &   | & \mathit{true} \\
	    &   | & \mathit{false}
	  \end{array}
	\end{align*}
    
    Additionally, if a term $\ite{v}{B}{B}$ occurs (both $B_t$ and $B_e$ are the same), it is replaced with $B$.
\end{definition}

The semantics of ITE normal form are defined under an assignment.  An assignment $\assignment$ maps each variable $\variable$ to a either the top element $\mathbf{1}$ or the bottom element $\mathbf{0}$.  We use the notation $v \mapsto \mathbf{1} \in \assignment$ to say that under the assignment $\assignment$, the variable $\variable$ has the value $\mathbf{1}$.  We use the judgment $\assignment \vdash B \Downarrow r$ to say that under the assignment $\assignment$, the formula $B$ evaluates to the value $r$, where $r$ is either $\mathbf{1}$ or $\mathbf{0}$.  The semantics follow:

\begin{mathpar}
\inferrule[I-ITE-T]
  {
   \variable \mapsto \mathbf{1} \in \assignment \\
   \assignment \vdash B_t \Downarrow r
  }
  {\assignment \vdash \ite{v}{B_t}{B_e} \Downarrow r}
\and
\inferrule[I-ITE-F]
  {
    \variable \mapsto \mathbf{0} \in \assignment \\
    \assignment \vdash B_e \Downarrow r
  }
  {\assignment \vdash \ite{v}{B_t}{B_e} \Downarrow r}
\\
\inferrule[I-True]
  { }
  {\assignment \vdash \mathit{true} \Downarrow \mathbf{1}}
\and
\inferrule[I-False]
  { }
  {\assignment \vdash \mathit{false} \Downarrow \mathbf{0}}
\end{mathpar}

A formula expressed in if-then-else normal form is also a binary decision diagram.  However, the most commonly used form of binary decision diagrams is assumed to be reduced and ordered.

\begin{definition}[Reduced Ordered Binary Decision Diagram]
    \label{def:robdd}
    A reduced ordered binary decision diagram (referred to as a BDD) is a formula written in if-then-else normal form with the following two additional restrictions:
    \begin{enumerate}
        \item There is a total order $\prec$ among variables $\variable$.  If $\variable_1 \prec \variable_2$ and both variables occur in the formula, then in the evaluation $\variable_1$ must be used before $\variable_2$.
        \item Sharing of formulas is mandated, so that if the same formula $B$ occurs more than once in the same formula, it is shared.
    \end{enumerate}
\end{definition}

The semantics of BDDs is the same as for if-then-else normal form.

These restrictions give BDDs canonicity and efficiency.  For a given ordering, there is only one BDD that represents a given formula.  Additionally, because of the sharing mandate, operations that are applied over a whole BDD often need only be applied once to each physically unique formula and thus sharing reduces work for many algorithms.

In addition to canonicity and efficiency, BDDs support quantifier elimination.  Both exponential and universal quantifiers can be eliminated from formulas with reasonable efficiency.  This is why BDDs are often preferred for solving QBF problems~\cite{pan2003optimizing,benedetti2005skizzo}.

BDDs are represented as a directed, acyclic graph where each vertex is an $\itef{}$ function.  The vertex is labeled with the variable that is being used in the if-then-else condition.  The two outgoing edges represent the $\mathit{else}$ case on the left with a solid line and the $\mathit{then}$ case on the right with a dashed line.  A $\mathit{false}$ is represented with $\bot$ and a $\mathit{true}$ is represented with $\top$ inside a vertex with no outgoing edges.

\begin{example}[BDDs representation of logic]

Consider the following formula $f$:
\begin{align*}
 f = v_1 \wedge \neg v_3 \vee \neg v_2 \wedge \neg v_3 
\end{align*}

Representing this formula in if-then-else normal form gives the following structure, assuming
the ordering $v_1 \prec v_2 \prec v_3$ was chosen.

\begin{align*}
\ite{v_1}
  {\ite{v_3}
     {\mathit{false}}
     {\mathit{true}}}
  {\ite{v_2}
     {\mathit{false}}
     {\ite{v_3}
        {\mathit{false}}
        {\mathit{true}}}}
\end{align*}

The BDD of the same formula is shown in Figure~\ref{fig:bdd-ex}.  It is the same as the if-then-else normal form except that it exploits sharing.  Note that $\ite{v_3}{\mathit{false}}{\mathit{true}}$ occurs twice in the formula and the equivalent node $v_3$ only occurs once in the BDD.  The two incoming arrows to $v_3$ indicate that sharing has improved the efficiency of the representation.

\begin{figure}[htb]
\centering
\begin{tikzpicture}
  [->,thick,
   every node/.style={draw,circle}]
  \node (v1) at (0,0) {$v_1$};
  \node (v2) at (-1,-1) {$v_2$};
  \node (v3) at (0,-3) {$v_3$};
  \node (t) at (-1,-4) {$\top$};
  \node (b2) at (1,-4) {$\bot$};
  \draw (v1) to[out=225,in=45] (v2);
  \draw[dashed] (v1) to[out=315,in=45] (v3);
  \draw[dashed] (v2) to[out=315,in=45] (b2);
  \draw (v2) to[out=225,in=135] (v3);
  \draw (v3) to[out=225,in=45] (t);
  \draw[dashed] (v3) to[out=315,in=135] (b2);
\end{tikzpicture}
\caption{BDD representation of the formula
$v_1 \wedge \neg v_3 \vee \neg v_2 \wedge \neg v_3$}
\label{fig:bdd-ex}
\end{figure}
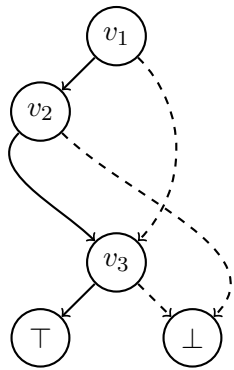

\end{example}

Not all is rosy for BDDs, however.  They are limited by the total ordering on the variables.  One ordering may be exponentially more efficient than another.  This means that efficient use of BDDs requires a good ordering.  Fortunately, achieving perfect efficiency is not required and there are many algorithms for idetifiying good orderings.

\subsection{Sets as Binary Decision Diagrams}
Since BDDs are well suited for representing Boolean algebras and sets are Boolean algebras, a BDD can be used as a set abstraction.  However, in practice, this is often too coarse of an abstraction.  Set operations often utilize the values stored within the sets.  For example, selecting an element from a set or computing the comprehension of a set.  These kind of operations are not trivially represented by a BDD as it only is suitable for reasoning about sets \emph{as-a-whole}.  Any reasoning about contents is lost.

In the remainder of this paper we present two things:  (1) we present the basic as-a-whole abstraction of sets using BDDs and give a couple of examples of where this type of abstraction may be useful; and  (2) we present extensions to the basic as-a-whole abstraction to support some amount of content reasoning.  Specifically, we focus on a prefix string abstraction to reason about contents of sets that are strings prefixed by string constants.

\section{Abstracting Set Constraints with BDDs}
\label{sec:abstraction}

In this section, we present an abstraction for symbolic, as-a-whole sets.  Symbolic, as-a-whole sets abstract away the individual constituents of sets and focus entirely upon the relationships between sets.  For example, symbolic, as-a-whole sets would be able to precisely abstract constraint $ A \subseteq B$, but would not be able to precisely abstract the constraint $A = {1, 2, 5}$.

In this section, we will use the variables $A$, $B$, and $C$ to represent set variables.  These symbols are members of the $\variables$ set.  For examples, we will assume that the sets we are abstracting contain integers $\mathbb{Z}$, but for the formalization we do not define $\values$ as the sets can contain any non-empty type of values.

The concrete program state that we will be abstracting is a valuation, which assigns the set value to each set symbol.  A valuation $\valuation$ is a member of this concrete program state $\conc$, which is defined as such:
\begin{align*}
\valuation \in \conc = \variables \rightarrow \powerset{\values}
\end{align*}
This means that in each $\valuation$ each variable $\variable \in \variables$ maps to a set of values.

\begin{definition}[BDD-based, symbolic, as-a-whole set abstraction]
    A BDD-based, symbolic, as-a-whole set abstraction is a binary decision diagram with the normal syntax:
    \begin{align*}
      \begin{array}{rcl}
        BDD \ni B & ::= & \ite{\variable}{B_t}{B_e} \\
        &   | & \mathit{true} \\
        &   | & \mathit{false}
      \end{array}
    \end{align*}
    Additionally, it has the orderedness and compactness restrictions given in Definition~\ref{def:robdd}.
\end{definition}

The concretization of a BDD into a valuation is given in two parts.  First we define the concretization $\gamma$, selectively validates elements returned by the second part.  The second part $\gamma_S$ constructs a concretization of the BDD augmented with a validation set $S$.

The functions $\gamma$ and $\gamma_S$ have the following types.
\begin{align*}
  \gamma &: BDD \rightarrow \powerset{ \conc } \\
  \gamma_S &: BDD \rightarrow \powerset{ \conc \times \powerset{\values}}
\end{align*}
The $\gamma$ function takes a BDD and returns a set of valuation functions.  To do this, it calls the $\gamma_S$ function, which returns a set of candidate valuations paired with a validation set.  If the valuation set is equal to the universe, that is the set of all values $\values$, that valuation function has been validated and can be included in the concretization.  The definition of $\gamma$ follows.

\begin{align*}
\gamma(B) &= \Set{ \valuation | (\valuation, \values) \in \gamma_S(B)}
\end{align*}

The construction of the valuation set follows from the Boolean algebra that BDDs are constructed from.  We can see this construction in the definition of $\gamma_S$.
\begin{align*}
\gamma_S(\mathit{false}) &= \Set{ (\valuation,\emptyset) | \valuation \in \conc} \\
\gamma_S(\mathit{true})  &= \Set{ (\valuation,\values) | \valuation \in \conc} \\
\gamma_S(\ite{\variable}{B_t}{B_e}) &= \Set{ (\valuation,S) |
    \begin{array}{l}
      (\valuation, S_t) \in \gamma_S(B_t) \wedge (\valuation, S_e) \in \gamma_S(B_e) \\
      {} \wedge S = (\comp{\valuation(\variable)} \cup S_t) \cap (\valuation(\variable) \cup S_e)
    \end{array} 
}
\end{align*}

The $\gamma_S$ function is defined in three parts.  One for each syntactic class of a BDD.  Note that the additional orderedness and compactness restrictions do not affect the concretization, only the efficiency and canonicity of the representation.  The first two classes reveal the nature of the validation set.  The application of $\gamma_S$ to $\mathit{false}$ gives the set of all possible valuations paired with the validation set $\emptyset$.  Since the $\emptyset$ validation set is never equal to $\values$, none of these valuations will be validated.  The converse is the $\mathit{true}$ case, where $\values$ is by definition equal to $\values$, so all possible valuations are validated.

The third syntactic class, which handles the $\itef{}$ function, performs the conditional operation on the validation set.  It computes the validation sets for both the $\mathit{then}$ ($S_t$) branch and the $\mathit{else}$ ($S_e$) branch.  The new validation set can be computed by looking up the set for the current variable $\variable$ in the valuation.

The reason for this formula comes from the correspondence between set algebra and Boolean algebra.  The operation $\ite{\variable}{B_t}{B_e}$ has the following definition in Boolean algebra:
\begin{align*}
  \variable \rightarrow B_t \wedge \neg \variable \rightarrow B_e
\end{align*}
Assuming the correspondence given between $\wedge$ and $\cap$, $\vee$ and $\cup$, and $\neg$ and $\comp{}$, the formula for the computation of the new validation set follows directly.

\subsection{Domain Operations}
\label{sec:domain-ops}

The domain operations for this set domain are derived directly from the equivalent BDD operations.  Typical BDD implementations provide at least the basic $\wedge$, $\vee$, and $\neg$ operations, along with universal and existential quantification.  The domain operations can be derived from those.

\paragraph{Constructing Expressions} Because of the lack of support for content reasoning, set expressions are limited to the following:
\begin{align*}
  E & ::= \emptyset \ | \ \values \ | \ A \ | \ E \cup E \ | \ E \cap E \ | \ E \uplus E \ | \ E \setminus E \ | \ \comp{E} \\
\end{align*}
This language incorporates all of the symbolic expressions for sets, including union, intersection, disjoint union, set difference and set complement.

To construct the binary decision diagrams that correspond to these expressions, we use a translation function $\tre{}$ that converts a set expression into a pair of binary decision diagrams.  This function is shown in Figure~\ref{fig:tr-expr}.  The first resulting BDD of this function is the translated expression and the second resulting BDD represents side constraints on that expression.  These side constraints are necessary to translate the disjoint union expression, which has a side constraint that the sets being unioned are disjoint.  All other operations simply pass along the constraints conjoining them in the BDD.

\begin{figure}[htbp]
\begin{align*}
  &\tre{\emptyset} = (\mathit{false}, \mathit{true}) & &\tre{E_1 \uplus E_2} = \\
  &\tre{\values} = (\mathit{true}, \mathit{true})    & & \qquad \letin{(e_1,c_1) = \tre{E_1}} \\
  &\tre{A} = (A, \mathit{true})                      & & \qquad \letin{(e_2,c_2) = \tre{E_2}} \\
  &\tre{E_1 \cup E_2} =                              & & \qquad (e_1 \vee e_2, c_1 \wedge c_2 \wedge \neg (e_1 \wedge e_2)) \\
  & \qquad \letin{(e_1,c_1) = \tre{E_1}}             & &\tre{E_1 \setminus E_2} = \\
  & \qquad \letin{(e_2,c_2) = \tre{E_2}}             & & \qquad \letin{(e_1,c_1) = \tre{E_1}} \\
  & \qquad (e_1 \vee e_2, c_1 \wedge c_2)            & & \qquad \letin{(e_2,c_2) = \tre{E_2}} \\
  &\tre{E_1 \cap E_2} =                              & & \qquad (e_1 \wedge \neg e_2, c_1 \wedge c_2) \\
  & \qquad \letin{(e_1,c_1) = \tre{E_1}}             & &\tre{\comp{E}} = \\
  & \qquad \letin{(e_2,c_2) = \tre{E_2}}             & & \qquad \letin{(e,c) = \tre{E}} \\
  & \qquad (e_1 \wedge e_2, c_1 \wedge c_2)          & & \qquad (\neg e, c) \\
\end{align*}
\caption{The translation function $\tre{}$ that converts a set expression into a pair of BDDs.  The first BDD represents the expression and the second BDD represents side constraints on that expression.}
\label{fig:tr-expr}
\end{figure}

Figure~\ref{fig:tr-expr} shows that the translation is the literal replacement of set operations with the corresponding BDD operation.  Of course, operations that deal with individual values will have to be abstracted into this language.  This means that expressions like singleton sets (for example $\{1\}$) have to be abstracted by a set symbol (for example $A$).  Other operations such as comprehensions (e.g.  $\Set{ x \in A | p(x) }$) can be abstracted by introducing a symbol and constraining that symbol (e.g. $B$ in the expression with the side constraint $B \subseteq A$).  The exact form of this abstraction is unspecified here.  The mere requirement is that the set expression be translated into this language to that it can be precisely translated via $\tre{}$ to its BDD equivalent.

\paragraph{Constructing Constraints} Constructing constraints from expressions requires a similar language restriction:
\begin{align*}
  K & ::= \mathit{true} \ | \ \mathit{false} \ | \ E \subseteq E \ | \ E = E \ | \ K \wedge K \ | \ K \vee K
\end{align*}
The language supports several commonly used set constraints including (non-strict) subset constraints between two set expressions and equality between two set expressions.  Additionally, it supports several standard Boolean combinators including conjunction and (somewhat uniquely for an abstract domain) disjunction.  Other operations are unsupported because they cannot be implemented solely with BDDs using the mechanism presented here.

The translation of constraints into BDDs does two things.  First, it actually constructs the constraints out of the constituent expressions or constraints.  Second, it merges the side-constraint BDDs that are produced by the translation of expressions into the constraints so that there is only a single BDD as a result.  The resulting BDD compactly represents the set expressions and set constraints together.

The definition of the translation function $\trk{}$ is shown in Figure~\ref{fig:tr-cons}.  It translate Boolean constraints directly into their BDD counterparts.  The subset constraint uses a Boolean implication ($\neg e_1 \vee e_2$) to merge the two expression BDDs.  The side constraints are conjoined to the constraint that utilized the expressions producing those constraints.  The equality constraint is similar to the subset constraint except that it uses a bi-implication instead of the single implication.

\begin{figure}[htbp]
    \begin{align*}
    &\trk{\mathit{true}} = \mathit{true}                                          & &\trk{K_1 \wedge K_2} = \trk{K_1} \wedge \trk{K_2} \\
    &\trk{\mathit{false}} = \mathit{false}                                        & &\trk{K_1 \vee K_2} = \trk{K_1} \vee \trk{K_2} \\
    &\trk{E_1 \subseteq E_2} =                                                    & &\trk{E_1 = E_2} =\\
    & \qquad \letin{(e_1,c_1) = \tre{E_1}}                                        & & \qquad \letin{(e_1,c_1) = \tre{E_1}}\\
    & \qquad \letin{(e_2,c_2) = \tre{E_2}}                                        & & \qquad \letin{(e_2,c_2) = \tre{E_2}}\\
    & \qquad (\neg e_1 \vee e_2) \wedge c_1 \wedge c_2                            & & \qquad (\neg e_1 \vee e_2) \wedge (\neg e_2 \vee e_1) \wedge c_1 \wedge c_2
    \end{align*}
    \caption{The translation function $\trk{}$ that converts set constraints into a BDD}
    \label{fig:tr-cons}
\end{figure}

Using these constraint forms, it is possible to implement the abstraction and/or the constrain domain operations.  These are useful for defining transfer functions for the program.

\paragraph{Join and Widening}  The join and widening operations are trivial.  Since the constraint language supports disjunction, the disjunction of the two BDDs gives a precise join:
\begin{align*}
  B_1 \sqcup B_2 = B_1 \vee B_2
\end{align*}
Critically, because this is represented using a Boolean algebra, the resulting lattice is finite height (for a fixed set of variables $\variables$).  This means that the join is also a suitable widening.  It may not be an optimal widening as the lattice has an exponential height and may take a long time to converge.  This is the challenge in BDD-based forward reachability and suggests that there may be ways of improving this widening operator using model checking techniques.

\paragraph{Containment}  The containment operation is also trivial.  It relies upon the implication ordering in the Boolean algebra lattice:
\begin{align*}
  B_1 \sqsubseteq B_2 = B_1 \rightarrow B_2
\end{align*}
The implementation of this is not quite as trivial, however.  This is because there is implicit universal quantification in the above formula.  Implication must be valid.  As a result it is possible that, because BDDs support universal quantification, the following could be implemented:
\begin{align*}
  \forall \bar{\variable}. \neg B_1 \vee B_2
\end{align*}
where $\forall \bar{\variable}.$ universally quantifies over each variable $\variable$ in $\variables$.  However, it is much more efficient to check the satisfiability of the negation:
\begin{align*}
  \mathrm{SAT}(B_1 \wedge \neg V_2)
\end{align*}
If the formula is unsatisfiable, its negation must be valid.

\paragraph{Projection} Projecting out variables is the primary reason BDDs are preferable for this application over SAT solvers.  BDDs support existential quantification directly and consequently it can be used to implement projection.  For example to project out the variable $\variable$ from the set domain instance $B$, the following BDD operation can be used:
\begin{align*}
  \exists \variable. B
\end{align*}

\section{Set Contents}
\label{sec:contents}

The abstraction presented in Section~\ref{sec:abstraction} does not permit reasoning about any contents of sets.  It is strictly an as-a-whole abstraction.  However, it can be adapted for varying amounts of content reasoning by implementing query-based reductions.  Query-based reductions utilize an external domain for keeping track of possible contents of sets and then use queries on the BDD to drive reductions in that domain.

\textcolor{blue}{(This section will be fleshed out in a future version)}

\section{Related Work}
\label{sec:related}

There exist two set abstract domains that focus on as-a-whole reasoning: QUIC graphs \cite{DBLP:conf/ecoop/CoxCS13} and FixBag \cite{DBLP:conf/cav/PhamTTC11}.  Both of these systems do not use a canonical representation like BDDs.  They use a proof system along with a heuristic-guided saturation/proof search.  This methodology lends itself to efficiently keeping track of content information about sets, but it is much less ideal for efficiently doing as-a-whole reasoning.  While both the BDD-based approach and the QUIC graphs/FixBag approach have exponential worst case, the QUIC graph and FixBag approach often encounter that worst case because the saturation technique attempts to enumerate all $O(2^n)$ possibilities.  The BDD-based approach often does not encounter this problem because it uses the ordering and the sharing to often eliminate the exponential cost.

There is no good comparison for the BDD-based set domain extended with content reasoning with QUIC graph/FixBag.  The reason is that BDD-based reasoning is heavily focused on precisely performing as-a-whole reasoning, while sacrificing precision (or slowing down) in the content reasoning.  Conversely, QUIC graphs/FixBag sacrifice as-a-whole precision and performance to get better content reasoning.  Different applications may have different needs.

The use of BDDs for model-checking-style verification is well documented~\cite{DBLP:conf/spin/Clarke08}.  BDDs were used to help solve the state explosion problem by symbolically representing many states implicitly~\cite{DBLP:conf/lics/BurchCMDH90,McMillan:thesis:1992}.  Of course, the complexity of these approaches compares with the complexity of abstract interpretation using BDDs to represent sets.  This is because if sets are used extensively, the set structure will end up capturing much of the control flow.  This results in the BDD needing to solve similar problems to model checking, which implicitly represents the control flow in the logic along with the data.  Of course the use of BDDs for symbolic model checking does not take advantage of the fact that they can represent things other than \textit{true} or \textit{false} values.

The use of BDDs for abstract domains is more recent.  They have been primarily used for logico-numeric abstraction in \textsc{BddApron}~\cite{bddapron}.  \textsc{BddApron} combines the \textsc{Apron} numeric abstract domain library~\cite{DBLP:conf/cav/JeannetM09} with BDDs to efficiently support disjunction.  The idea is to the the BDD to represent the control flow, but to use the \textsc{Apron} domains for numeric reasoning.  Aside from the fact that it is intended to capture control flow, this is similar to what BDD-based sets does with reductions.  The set abstraction restricts another abstraction that reasons about values.

\section{Conclusion}
\label{sec:conclusion}

This paper describes a new method for abstracting states of set-manipulating programs.  The domain is built upon the logical foundation of binary decision diagrams and utilizes the fact that binary decision diagrams can represent any Boolean logic, not just the standard $\mathbf{0}-\mathbf{1}$ logic.  By using the well-engineered, well-designed data structures for BDDs, we have found that when as-a-whole reasoning is needed, it is generally more efficient and more precise to use BDDs than other proof-based methods.

Of course, the sacrifice that was made was in terms of content reasoning.  The BDDs do not provide a native way for reasoning about the specific contents of sets.  This can be remedied somewhat by using a value abstraction and performing query-based reductions with the BDDs.  However this space remains to be explored more thoroughly.  The query-based approach introduces significant run-time overhead if applied thoroughly.  Either good heuristics should be employed or a new hybrid approach should be developed.  What works best remains to be seen.

\bibliographystyle{alpha} 
\bibliography{cox}    

\newcommand{\etalchar}[1]{$^{#1}$}
\begin{thebibliography}{BCM{\etalchar{+}}90}

\bibitem[BCM{\etalchar{+}}90]{DBLP:conf/lics/BurchCMDH90}
Jerry~R. Burch, Edmund~M. Clarke, Kenneth~L. McMillan, David~L. Dill, and L.~J.
  Hwang.
\newblock Symbolic model checking: 10{\^{}}20 states and beyond.
\newblock In {\em LICS}, pages 428--439, 1990.

\bibitem[Ben05]{benedetti2005skizzo}
Marco Benedetti.
\newblock skizzo: a suite to evaluate and certify qbfs.
\newblock In {\em Automated Deduction--CADE-20}, pages 369--376. Springer,
  2005.

\bibitem[CC77]{DBLP:conf/popl/CousotC77}
Patrick Cousot and Radhia Cousot.
\newblock Abstract interpretation: A unified lattice model for static analysis
  of programs by construction or approximation of fixpoints.
\newblock In {\em POPL}, pages 238--252, 1977.

\bibitem[CCR14]{HOO}
Arlen Cox, Bor-Yuh~Evan Chang, and Xavier Rival.
\newblock Automatic analysis of open objects in dynamic language programs.
\newblock In {\em SAS}, 2014.

\bibitem[CCS13]{DBLP:conf/ecoop/CoxCS13}
Arlen Cox, Bor-Yuh~Evan Chang, and Sriram Sankaranarayanan.
\newblock {QUIC} graphs: Relational invariant generation for containers.
\newblock In {\em ECOOP}, pages 401--425, 2013.

\bibitem[Cla08]{DBLP:conf/spin/Clarke08}
Edmund~M. Clarke.
\newblock The birth of model checking.
\newblock In {\em 25 Years of Model Checking - History, Achievements,
  Perspectives}, pages 1--26, 2008.

\bibitem[Jea09]{bddapron}
Bertrand Jeannet.
\newblock Bddapron: A logico-numerical abstract domain library.
\newblock
  \url{http://pop-art.inrialpes.fr/\~{}bjeannet/bjeannet-forge/bddapron/},
  2009.

\bibitem[JM09]{DBLP:conf/cav/JeannetM09}
Bertrand Jeannet and Antoine Min{\'e}.
\newblock Apron: A library of numerical abstract domains for static analysis.
\newblock In {\em CAV}, pages 661--667, 2009.

\bibitem[Kun07]{Kuncak:Thesis:2007}
Viktor Kuncak.
\newblock {\em Modular Data Structure Verification}.
\newblock PhD thesis, EECS Department, Massachusetts Institute of Technology,
  February 2007.

\bibitem[McM92]{McMillan:thesis:1992}
Kenneth~L. McMillan.
\newblock {\em Symbolic Model Checking: An approach to the state explosion
  problem}.
\newblock PhD thesis, Carnegie Mellon University, 1992.

\bibitem[PTTC11]{DBLP:conf/cav/PhamTTC11}
Tuan-Hung Pham, Minh-Thai Trinh, Anh-Hoang Truong, and Wei-Ngan Chin.
\newblock Fixbag: A fixpoint calculator for quantified bag constraints.
\newblock In {\em CAV}, pages 656--662, 2011.

\bibitem[PV03]{pan2003optimizing}
Guoqiang Pan and Moshe~Y Vardi.
\newblock Optimizing a bdd-based modal solver.
\newblock In {\em Automated Deduction--CADE-19}, pages 75--89. Springer, 2003.

\end{thebibliography}

\end{document}